\documentclass[runningheads]{llncs}

\usepackage[utf8]{inputenc}
\usepackage[english]{babel}
\usepackage{graphicx}
\usepackage{amssymb}
\usepackage{mathtools}
\usepackage{here}
\usepackage{tabularx}
\usepackage{amsmath}
\usepackage{cite}

\newcommand{\NNR}{{\sc NNReach}}
\newcommand{\VIP}{{\sc VIP}}
\newcommand{\NE}{{\sc NE}}
\newcommand{\R}{{\mathbb{R}}}
\newcommand{\Q}{{\mathbb{Q}}}

\DeclareMathOperator{\CSP}{CSP}

\begin{document}

\begin{Large}
Preprint "Complexity of Reachability Problems in Neural Networks", May 2023, Adrian Wurm\\

\end{Large}

\title{Complexity of Reachability Problems in Neural Networks}
\author{Adrian Wurm }

\authorrunning{A. Wurm}

\institute{BTU Cottbus-Senftenberg, Lehrstuhl Theoretische Informatik,\\ Platz der Deutschen Einheit 1, 03046 Cottbus, Germany,
\url{https://www.b-tu.de/}
\email{wurm@b-tu.de}}

\maketitle              
%


\textbf{Abstract:} In this paper we investigate formal verification problems for Neural Network computations. Various reachability problems will be in the focus, such as: Given symbolic specifications of allowed inputs and outputs in
form of  Linear Programming instances, one question is whether valid inputs exist such that the given network computes a valid output? Does this property hold for all valid inputs? The former question's complexity has been investigated recently in \cite{lange} by S\"alzer and Lange for nets using the Rectified Linear Unit and the identity function as their activation functions.
We complement their achievements by showing that the problem is NP-complete for piecewise linear functions with rational coefficients that are not linear, NP-hard for almost all suitable activation functions including non-linear ones that are continuous on an interval, complete for the Existential Theory of the Reals $\exists \mathbb R$ for every non-linear polynomial and $\exists \mathbb R$-hard for the exponential function and various sigmoidal functions. For the completeness results, linking the verification tasks 
with the theory of Constraint Satisfaction Problems
turns out helpful.

\section{Introduction}

Given the huge success of utilizing (Deep) Neural Networks, (D)NN for short, in the last decade, such nets are nowadays widely used in all kind of data processing, including tasks of varying difficulty.
There is a wide range of applications, the following exemplary references (mostly taken from \cite{lange}) just 
collect non-exclusively some areas 
for further reading: Image recognition \cite{Krizhevsky}, natural language processing \cite{Hinton}, 
autonomous driving \cite{Grigorescu},
applications in medicine \cite{Litjens}, and prediction of stock markets \cite{Dixon}, just to mention a few. 
Khan et al. \cite{Khan} provide a survey of such applications, a mathematically oriented textbook
concerning structural issues related to Deep Neural Networks is provided by \cite{Calin}. Among the many different aspects of areas where the use of Neural Networks seems appropriate,
some also involve safety-critical systems like autonomous driving 
or power grid management. In such a setting, when security issues become important, aspects of certification come into play \cite{Huang}.

In the present paper we are interested in studying certain verification problems for
NNs in form of particular reachability problems.
Starting point is the work by S\"alzer and Lange \cite{lange} being based on \cite{Katz,Ruan}. The authors of these papers analyze the computational
complexity of one particular such verification task. It deals with a reachability problem 
in networks using the Rectified Linear Unit
together with the identity function as its activation function. In such a net, specifications describe
the set of valid inputs and outputs in form of two Linear Programming instances. The
question then is to decide whether a valid input exists such that the network's result
on that input is a valid output, i.e., whether the set of valid outputs is reachable from
that of valid inputs. In the above references the problem is shown to be NP-complete, even for one hidden layer and output dimension one, or some restricted set of weights being used \cite{lange}. 
Note that the network in principle is allowed to compute with real numbers, so the valid inputs we are looking for belong to some space $\mathbb R^n$, but the network itself is specified by
its discrete structure and rational weights defining the linear combinations computed by its single neurons.

Obviously, one can consider a huge variety of networks created by changing the underlying activation functions. There are of course many activations frequently used in NN frameworks, and in addition we could extend reachability questions to nets using all kinds of activation. One issue to be discussed is the computational model in
which one argues. If, for example, the typical sigmoid activation $f(x) = 1/(1 + e^{-x})$ is used, it has to be specified in which sense it is computed by the net: For example exactly or approximately, and at which costs these
operations are being performed.


In the present work we study the reachability problem for commonly used activation functions
and show that for most of them it will be complete either in (classical) NP or 
in the presumably larger class $\exists\mathbb R$, which captures the so-called existential theory of the reals.
Our main results are as follows:
The reachability problem is in P for linear activations, in NP for semilinear activations, NP-hard for all non-linear activations that are continuous on an interval, and ETR-hard for several commonly used activations such as arctan and the exponential function. These results imply, for example, NP-completeness for frequently used
activations such as (Leaky) ReLU, Heaviside and Signum.


A most helpful tool for establishing these results is linking the problems
under consideration to the area of constraint satisfaction problem CSP and
known complexity results for special instances of the latter.
This connection will provide us with a classification
of a vast set of activation functions in the complexity classes between P and $\exists\mathbb R$.
We also consider a variant of the reachability problem asking whether for all
valid inputs the computed output is necessarily valid and establish several complexity results as well.

The paper is organized as follows:
In Section \ref{Section:preliminaries} we collect basic notions, recall the definition of feedforward neural nets 
as used in this paper as well as useful facts about Constraint Satisfaction Problems.
Section \ref{Section:reachability} studies various activation functions and their impact on the complexity of reachability problems. We show that the reachability problem is basically the same as the CSP containing the graphs of the activation functions together with relations necessary to express linear programming instances. We show that adding the identity as activation does not change the complexity of the reachability problem in several cases, for example when either ReLU is used as activation or if a network connection is allowed to skip a layer. We show that the problem is NP-hard for every sensible non-linear activation and finally discuss problems that are hard or complete for $\exists\mathbb R$. The paper ends with some open questions. Lacking proofs are given in the Appendix.

\section{Preliminaries and Network Reachability Problems}\label{Section:preliminaries}

We start by defining the problems we are interested in; here,
we follow the definitions and notions of \cite{lange} for everything related to 
neural networks. The networks considered are exclusively feedforward. In their
most general form, they can process real numbers and contain rational weights. 

\begin{definition} A (feedforward) neural network N is a layered graph that represents a
function of format $\mathbb R^n \rightarrow \mathbb R^m$, for some $n,m \in \mathbb N$.
The first layer with label $\ell = 0$ is called the \emph{input layer} and consists of $n$ nodes called \emph{input nodes}. 
The input value $x_i$ of the $i$-th node is also taken as its output $y_{0i}  := x_i$.
A layer $1 \leq \ell \leq L - 2$ is called \emph{hidden} and consists of $k(\ell)$ nodes called \emph{computation nodes}. The $i$-th node of layer
$\ell$ computes the output $y_{{\ell}i} = \sigma_{{\ell}i} (\sum\limits_j c_{ji}^{({\ell}-1)}y_{({\ell}-1)j} + b_{{\ell}i})$. 
Here, the $\sigma_{{\ell}i}$ are (typically nonlinear)  \emph{activation} functions (to be specified later on) and the
sum runs over all output neurons of the previous layer. The $c^{({\ell}-1)}_{ji}$ are real constants which are called \emph{weights}, and $b_{{\ell}i}$ is a real constant  called \emph{bias}.
The outputs of all nodes of layer $\ell$ combined
gives the output $(y_{\ell 0}, . . . , y_{\ell(k-1)})$ of the hidden layer.
The final layer $L - 1$ is called \emph{output layer} and consists of $m$ nodes called \emph{output nodes}.
The $i$-th node computes an output $y_{(L-1)i}$ in the same way as a node in a hidden
layer. The output $(y_{(L-1)0}, . . . , y_{(L-1)(m-1)})$ of the output layer is considered
the output $N(x)$ of the network $N$.
\end{definition}

Note that above we allow several different activation functions in a single network. This basically is because for some results technically the identity is necessary as a second activation function 
beside the 'main' activation function used. 
We next recall from \cite{lange} the definition of the reachability problem \NNR.
Since we want to study its complexity in the Turing model, we restrict all weights and
biases in a NN to the rational numbers. The problem involves two Linear Programming LP instances
in a decision version, recall that such an instance consists of a system of (componentwise) linear
inequalities $A\cdot x \leq b$ for a rational matrix $A$ and vector $b$ of suitable dimensions.
The decision problem asks for the existence of a real solution vector $x.$

\begin{definition}\label{decision}
a) Let $F$ be a set of activation functions from $\mathbb{R}$ to $\mathbb{R}$.  
An instance of the \emph{reachability problem for neural networks} \NNR$(F)$  
consists of an $n \in \mathbb{N},$ a (feedforward) neural network $N$ with $n$ inputs and all its 
activation functions belonging to $F$, rational data as
weights and biases, and 
two instances of LP in decision version with rational data, 
one with the input variables of $N$ as variables, and the other 
with the output variables of $N$ as variables. 
These instances are also called \emph{input} and \emph{output specification}, respectively.
The problem is to decide if there exists an $x \in \mathbb R^n$ that satisfies the input specification such that the output $N(x)$ 
satisfies the output specification.

b) The problem \emph{verification of interval property} \VIP(F) consists of the same instances, except for the output specification being the open polyhedron, meaning the interior of the solution space. This is due to technical reasons that will later on simplify the reductions. The question is whether for all $x \in \mathbb{R}^n$ satisfying the input specification, $N(x)$ will satisfy the output specification (cf. \cite{Huang}).

As for \NNR, we denote by $(A,B,N)$ such an instance, assuming $n$ is obvious from the context.

c) Let $F=\{f_1,...,f_n\}$ be a set of activation functions. Then the Network Equivalence problem \NE$(F)$ is the decision problem whether two $F$-networks describe the same function or not.

d) The size of an instance is given as the sum of the (usual) bit-sizes of the two LP instances 
 and $T \cdot L$; here, $T$ denotes the number of neurons in the net $N$ and $L$ is the
maximal bit-size of any of the weights and biases.
\end{definition}

As usual for neural networks, we consider different choices for the activation functions used. Typical activation functions are $ReLU(x)=max\{0,x\}$, the Heaviside function or sigmoidal functions like $\sigma(x)=\frac1{1+e^{-x}}$. By technical reasons, in some situations the identity function $\sigma(x)=x$ is also allowed, Sälzer and Lange \cite{lange}, for example, examined \NNR$(id, ReLU)$. We name nodes according to their internal activation function, so we call nodes with activation function $\sigma(x)=x$ identity nodes and nodes with activation function $\sigma(x)=ReLU(x)$ ReLU-nodes etc. 
Note that the terminology of the LP-specifications has its origin in software verification.

\subsection{Basics on Constraint Satisfaction Problems CSP}

As we shall see, analyzing the complexity of the above reachability
problems is closely related to suitable questions in the
framework of Constraint Satisfaction Problems CSP. This is a well established area in
complexity, see for example the survey \cite{Carbonnel}. 
Here, we collect the basic notions and results necessary for our purposes.

Informally, a CSP deals with the question whether values from a set $A$ can be assigned
to a set of variables so that given 
conditions (constraints) hold. These conditions are taken from a 
set of relations over $A$ that, together with the set $A$, 
define the CSP. This can be formalized as follows:

\begin{definition} A (relational) \emph{signature} is a pair $\tau=(\mathbf{R},a),$ 
where $\mathbf{R}$ is a finite set of \emph{relation symbols} 
 and $a\colon\mathbf{R}\rightarrow \mathbb{N}$ is a function called 
the \emph{arity}.

 A (relational) $\tau$-\emph{structure} is a tuple $\mathcal{A}=(A,\mathbf{R}^\mathcal{A})$,  
where $A$ is a set called the \emph{domain} and $\mathbf{R}^\mathcal{A}$ is a set containing precisely one 
relation $R^\mathcal{A}\subseteq A^{a(R)}$ for each relation symbol $R\in\mathbf{R}.$





\end{definition}


An instance of a CSP over a given  $\tau$-structure is a conjunction of constraints, where a single constraint restricts a variable tuple to belong to a particular relation of the structure under a suitable assignment of values from the domain to the variables. For the entire instance one then asks for the existence of an assignment satisfying all its constraints.

\begin{definition}

Let $\tau$ be a signature and $\mathcal{A}$ a $\tau$-structure with domain $A$ and relations $\mathbf{R}.$ We always assume equality to be among the structure's relations. Let $X =\{x_1,x_2,\ldots\}$ be a countable set of variables.

a) A \emph{constraint} for $\mathcal{A}$ is an expression
$R(y_1,...,y_{a(R)})$, where $R\in\mathbf{R}, a(R)$ its arity and all $y_i\in X$. 
For $z \in A^{a(R)}$ we say that $R(z)$ is true over $\mathcal{A}$ iff $z \in R^\mathcal{A}.$

b)  A formula $\psi$ is called \emph{primitive positive} if it is of the form 
\[\exists x_{n+1},...,\exists x_{t} \colon \psi_1\land...\land\psi_k,\]
where each $\psi_i$ either is a constraint, $\top$ (true), or $\bot$ (false).

A formula with no free variables is a \emph{sentence}.

c) The decision problem $\CSP(\mathcal{A})$ is the following: Let $n \in \mathbb{N}$ and a primitive positive $\tau$-sentence over $X$, i.e., a finite collection of constraints involving variables $\{x_1,\ldots,x_n\} \subset X$ be given. 
The question then is, whether there exists an assignment $f : \{x_1,\ldots,x_n\} \to A$ for the variables such that each given constraint is true under the assignment $f$, i.e., all $R( f(y_1),...,f(y_{a(R)})) $ are true.

The size of an instance is $n+m,$ where $m$ denotes the number of constraints.

\end{definition}

\begin{example}[folklore]\label{lpffeaseq}
Consider as domain the real numbers $\mathbb{R}$, together with the binary order relation $\leq$,
the ternary relation $R_+$ defined via $R_+(x,y,z) \Leftrightarrow x+y=z$, and the
unary relation $R_{=1}(x) \Leftrightarrow x=1.$
Then $\CSP(\mathbb R; \leq,R_+,R_{=1})$ is polynomial time equivalent to the Linear Programming problem in feasibility form with rational input data. Reducing the former to the latter is obvious, for the reverse direction first multiply all inequalities with 
a sufficiently large natural number to obtain integer coefficients only. Now observe that any 
natural number $n$ can be expressed as (one component of) a solution of a set of constraints involving
$a=1$ and doubling a number via $ c = b + b.$ This way, the binary expansion of $n$ can be constructed with $O(\log{n})$
constraints. Apply this construction similarly to a variable $x$ of an instance of LP to obtain the term $n\cdot x$;
now adding as constraint the equation $nx =1$ similarly allows to express rational numbers as coefficients. 
Clearly the size of the resulting CSP instance is polynomially
bounded in the (bit-)size of the given LP instance.
Note that due to the theory of Linear Programming an instance with rational data has
the same answer, independently of whether the considered domain is $\R$ or $\Q.$
\end{example}

\begin{definition}
A relation $R$ is called \emph{primitive positive definable} (pp-definable) over  $\mathcal{A}$, iff it 
can be defined by a primitive positive formula $\psi$ over $\mathcal{A}$, i.e.,
\[R(x_1,...,x_n)\Leftrightarrow\exists x_{n+1},...,\exists x_{t} \colon \psi(x_1,\ldots, x_{t}).\]
\end{definition}

It was shown by Jeavons, Bulatov and Krokhin \cite{krokhin} that  $\CSP(\mathcal{A})$ and $\CSP(\mathcal{A'})$, where the latter structure arises from the former
by attaching finitely many relations being pp-definable over $\mathcal{A}$, are linear-time equivalent. The obvious idea of replacing every occurrence of the new relation suffices to prove the statement. This argument will be used below once in a while.

\begin{definition}
a) A set $R\subseteq\mathbb R^n$ is called \emph{semilinear}, iff it is a boolean combination of half-spaces\footnote{i.e., finite unions, intersections and complements of sets of the form $Ax\leq b$}.  

b) A set $R\subseteq\mathbb R^n$ is called \emph{essentially convex}, iff for any two points $x,y\in\mathbb R^n$ the intersection of the line segment $[x,y]$ contains only finitely many points that are not in $R$. If $R\subseteq\mathbb R^n$ is not essentially convex, any two points for which the property fails are called \emph{witnesses} for the set not being essentially convex.

\end{definition}

This gives us access to the following results of Bodirsky, Jonsson and von Oertzen:

\begin{theorem}\cite{Jonsson}\label{dual}
a) Let $R_1,...,R_n$ be semilinear relations. Then\\
$\CSP(\mathbb Q; \leq,R_+,R_{=1},R_1,...,R_n)$ is in P if $R_1,...,R_n$ are essentially convex and is NP-complete otherwise.

b) Let $R_1,...,R_n$ be relations such that at least one of them is not essentially convex witnessed by two rational points. Then $\CSP(\mathbb R; \leq,R_+,R_{=1},R_1,...,R_n)$ is NP-hard.\footnote{Note that we can not switch between the domains $\mathbb Q$ and $\mathbb R$ at will any more after dropping semilinearity with rational coefficients, for it could in this case change solvability.}
\end{theorem}



\section{Complexity Results for Reachability}\label{Section:reachability}
We shall now study the complexity of the reachability problem for various
sets of activation functions used by the neural network under consideration. Starting point
will be the result from \cite{Katz,lange} that \NNR$(id, ReLU)$ is NP-complete.
We analyze the problem for a larger repertoire of activation functions.
To do so, in a first step it will be very helpful to relate these problems to instances of certain CSP problems which can be attached to a network canonically.
This relation is made precise in the following theorem. The fact that input and output specifications are LP instances causes, that the structures below naturally
contain the relations $R_{=1}, R_{+},$ and ${\leq}.$ Further relations then will be
determined by the activation functions used.

\begin{theorem}\label{addfunction}
For any set of unary real functions $F=\{f_1,...,f_s\}$, interpreted as relations via their graphs, $\CSP(\mathbb R; \leq,R_+,R_{=1}, f_1,...,f_s)$ and\\
\NNR$(id,f_1,...,f_s)$ are linear-time equivalent.
\end{theorem}

\begin{proof}
We prove both directions explicitly for the case $s=1$, then the conclusion for $s>1$ is immediate.
For reducing \NNR$(id,f)$ to $\CSP(\mathbb R; \leq, R_+,R_{=1}, f)$, let $N$ be a network using $id$ and $f$ as activation functions. The weights and biases of $N$ are assumed to be rational numbers. The variable set of the CSP we construct contains one variable for each input and output node of $N$. For each node $v$ in a hidden layer we introduce two variables $v_{sum}$ and $v_f$. Note that according to Example \ref{lpffeaseq} any linear inequality with rational coefficients can be expressed as an instance of $\CSP(\mathbb R; \leq,R_+,R_{=1})$ of linear size. Thus, the input and output specifications of $N$ can be expressed as a set of constraints in $\CSP(\mathbb R; \leq,R_+,R_{=1})$ using the corresponding variables, which is of linear size with respect to the size of those specifications. For the nodes in the hidden layers, we proceed similarly. If node $v$ receives a linear sum $\sum\limits_{i=1}^k c_i\cdot u_i +b$ as its input, where $c_i,b$ are the rational weights and bias and the $u_i$ represent the outputs of the previous layer, then as in Example \ref{lpffeaseq} we add the constraint $v_{sum}=\sum\limits_{i=1}^k c_i\cdot u_{i,f} +b$ to the constructed instance. In case $v$ has $f$ as activation, we add the constraint $v_f=f(v_{sum})$ and if $v$ was an $id$-node we add the constraint $v_f=v_{sum}$. Obviously, the size of the CSP instance is linearly bounded in that of the given net. Moreover, \NNR$(id,f)$ is solvable for $N$ if and only if the above CSP has a solution.

For the reverse direction, we translate an instance of $\CSP(\mathbb R; \leq,R_+,R_{=1}, f)$ into an instance of \NNR$(id,f)$ with only one hidden layer. For each variable in the instance we introduce a node in the input layer and encode all constraints of the form $\leq,R_+$ and $R_{=1}$ into the input specification. For every constraint $y=f(x)$ 
we introduce a new $f$-node $\bar x$ in the hidden layer connected only to $x$ with bias 0 and weight 1. Next, we allocate an identity-node $\bar y$ in the hidden layer connected only to $y$ also with bias 0 and weight 1 for the connections. Finally, we require both nodes $\bar{x}$ and $\bar{y}$ to be equal by adding the equation  $\bar{x}=\bar{y}$ to the output-specification.  

It is obvious that both reductions can be performed inductively for all the functions 
in $F$, so the statement holds for the entire set.$\hfill\blacksquare$
\end{proof}
Note that the proof does not depend on formalizing the specifications as LP instances. It would similarly hold if the specifications would be given by (in-) equality systems involving polynomials and adding a relation for multiplication on the CSP-side. However, in this case checking feasibility of the specifications is already difficult, see below.

Before studying \NNR \ for different activations, we briefly discuss a more technical issue, namely the necessity of adding $id$ as activation. 

The above proof implies that using an injective activation allows to omit $id$,
if we drop the condition that the network has to be layered, 
meaning that a connection can skip layers:
\begin{lemma}
For $f$ injective, \NNR$(id,f)$ and \NNR$(f)$ are linear-time equivalent.
\end{lemma}

\begin{proof}

Given the proof of Theorem \ref{addfunction} it only remains to avoid $id$-nodes when reducing an instance of $\CSP(\mathbb R; \leq, R_+,R_{=1}, f)$
to one of \NNR$(f).$ Identity nodes were used to propagate the value of a node $y$ in order to include
a constraint $y= f(x)$ in the output specification.  Instead, if $f$ is injective one
can use a network node for $f(x)$ and one for $y$ and connect them with biases $0$ and weights
$1$ and $-1$, respectively, to an $f$-activation node computing $f(f(x) -y)$. Use another $f$-node
to compute $f(0).$ This is possible by demanding a further input node to have value $0$.
Finally, in the output specification we add the equality between these two nodes with values $f(f(x)-y)$ and $f(0)$; injectivity provides the equivalence of this condition with $f(x) = y.$  $\hfill\blacksquare$
\end{proof}

For example, for nets using sigmoidal activation functions identity activations are not necessary.

S\"alzer and Lange \cite{lange} asked whether for the NP-completeness result involving $id$ and $ReLU$ as activations 
one can avoid $id$  as activation. Though $ReLU$ is not injective, this in fact holds as well, even when we do not allow connections to skip layers:

\begin{proposition}\label{onlyReLU}
The following problems are linear-time equivalent:\begin{description}
\item[i)] $\CSP(\mathbb R; \leq,R_+,R_{=1}, ReLU)$,
\item[ii)] \NNR$(id,ReLU)$ and
\item[iii)] \NNR$(ReLU)$ with only one hidden layer.
\end{description}
As consequence, all three are NP-complete.
\end{proposition}

\begin{proof}
Given the NP-completeness of \NNR$(id,ReLU)$ and Theorem \ref{addfunction} above, it remains 
to reduce \NNR$(id,ReLU)$ to \NNR$(ReLU)$ in linear time. Towards this goal,
we show that an identity node can be replaced by two ReLU-nodes in the following way: In the neural net to be constructed use two copies of the identity node and let both have the same incoming and outgoing connections as the original node. Replace the identity map by the ReLU map in both and invert all incoming and outgoing weights as well as the bias in the second one. Delete the initial identity node. This does not change the computed function of the network, because
\begin{align*}
\sum\limits_{i=1}^na_ix_i+b&=max\{0,\sum\limits_{i=1}^na_ix_i+b\}+min\{0,\sum\limits_{i=1}^na_ix_i+b\}\\
&=max\{0,\sum\limits_{i=1}^na_ix_i+b\}-max\{0,-(\sum\limits_{i=1}^na_ix_i+b)\}\\
&=ReLU(\sum\limits_{i=1}^na_ix_i+b)-ReLU(-(\sum\limits_{i=1}^na_ix_i+b))\\
&=ReLU(\sum\limits_{i=1}^na_ix_i+b)-ReLU(\sum\limits_{i=1}^n(-a_i)x_i-b)
\end{align*}
Applying this to every node gives us at most twice as many nodes with at most four times as many connections, thus the construction runs in linear time.\hfill$\blacksquare$

\end{proof}



Theorem \ref{addfunction} enables us treating network
reachability complexity questions by using the rich fund of complexity results for CSP problems of various
types.

As an easy warm up, convince yourself that \NNR$(id)$ is by the previous theorem equivalent to $\CSP(\mathbb R; \leq, R_+,R_{=1},id)$ which in turn is equivalent to LP by Example \ref{lpffeaseq}, a problem well known to belong to P in Turing model complexity \cite{Karmarkar}.

In \cite{lange} it was shown that \NNR$(id,ReLU)$ is NP-complete, the link to CSPs however provides us with a much shorter proof. We will make use of Theorem \ref{dual} and apply Theorem \ref{addfunction}:




\begin{corollary}\label{dualNN}
Let $f_1,...,f_s$ be unary real functions. If their graphs $g_1,...,g_s$ are semilinear with rational coefficients, then \NNR$(id,f_1,...,f_s)$ is in P if and only if $g_1,...,g_s$, interpreted as binary relations, are essentially convex, and NP-complete otherwise. If at least one of the graphs $g_1,...,g_s$ is not essentially convex witnessed by two rational points, then \NNR$(id,f_1,...,f_s)$ is NP-hard.
\end{corollary}
Though Corollary \ref{dualNN} certainly is not that surprising from the CSP point of view, it gives an elegant way to argue about the complexity of reachability problems for neural networks. Of interest now is to investigate the latter for many more activation functions. NP-hardness is for example the case for $f(x)=n^x, n\in\mathbb{N} , n>1$, the witnesses are $f(0)=1$ and $f(1)=n$, but not necessarily for $f(x)=e^x$ for it lacks a second rational point.
Other activation functions that immediately give us NP-hardness this way are all non-linear rational functions (including polynomials) with rational coefficients, the square-root, all other rational roots, and the binary logarithm. Reasonable non-linear functions that lack rational points can be compressed to do so, such as $f(x)=sin(\frac x\pi)$ instead of $f(x)=sin(x)$.

Moreover, if we use the strong version that requires semilinearity, we get that \NNR$(id, f)$ is NP-complete for $f$ being either the absolute value, the floor or ceiling function on a bounded domain, the sign or the Heaviside function, piecewise linear functions such as
$f_\alpha(x) = \left\{
\begin{array}{ll}
-1 \ \ \ \ & if \ x \leq -\alpha \\
\frac x\alpha & if \ -\alpha<x<\alpha \\
1 & if \ x \geq \alpha \\
\end{array}
\right. $, 
the ReLU function, Leaky ReLU given via $R_\alpha(x)=
\left\{\begin{matrix}
x \ \ \ \ \ & if \  x\geq0\\
\alpha x \ \  & if  \ x<0
\end{matrix}\right. , \alpha\in\mathbb Q$ as well as all their scalar generalizations and any other non-trivial rational step function such as the indicator function on an interval.

We will now see a much more powerful version of the NP-hardness part of Corollary \ref{dualNN}:
\begin{theorem}\label{contin}
\NNR$(id,f)$ is NP-hard for any non-linear function $f:\mathbb{R}\rightarrow\mathbb{R}$ that is continuous on an interval $[a,b]\subseteq\mathbb R$, including $f(x)=\frac{x}{1+e^{-x}}$ and all sigmoidal functions such as $f(x)=\frac{1}{1+e^{-x}}$, $f(x)=\frac{x}{\sqrt{1+x^2}}$ and $f(x)=tanh(x)$ the hyperbolic tangent. 
\end{theorem}

\begin{proof}
By the previous Corollary, it suffices to show that with such a function $f$ we can pp-define a new function $\bar f$ that excludes an interval witnessed by two rational points. The construction of $\bar f$ proceeds in several steps.\\
First, by density of $\mathbb Q$ there exist $\bar a,\bar b\in[a,b]\cap\mathbb Q , \bar a<\bar b$ so that $f\vert_{[\bar a,\bar b]}$ is still non-linear and continuous, we apply a linear transformation on the argument of $f$ so that we can assume $\bar a=0$ and $\bar b=1$. This is pp-definable for rational $\bar a,\bar b$ and neither changes non-linearity nor continuity. There must exist $c,d\in[0,1]\cap\mathbb Q$, such that \[f(\frac{c+d}2)\neq\frac{f(c)+f(d)}2,\] for $f$ would otherwise fulfill Cauchy's functional equation and therefore be linear.

Apply again a linear transformation on the argument that maps $c$ to 0 and $d$ to 1 and call the resulting function $\hat f$. By construction,
\[\hat f(\frac{1}2)\neq\frac{\hat f(0)+\hat f(1)}2\]
Define the function $\bar f:\mathbb R\rightarrow \mathbb R$ by $\bar f(x)=\hat f(x)+\hat f(1-x)-\hat f(0)-\hat f(1)$, this is primitive positive, for addition, affine transformation and the constants 0 and 1 are pp-definable in $(\mathbb R;\leq,R_+,R_{=1})$. It also matches the requirements for Corollary \ref{dualNN} part 2: The rational points are $\bar f(0)=0$ and $\bar f(1)=0$ and the function must exclude an interval because 
\[\bar f(\frac12)=\hat f(\frac12)+\hat f(\frac12)-\hat f(0)-\hat f(1)=2(\hat f(\frac{1}2)-\frac{\hat f(0)+\hat f(1)}2)\neq0\]
 and by continuity.\hfill$\blacksquare$
\end{proof}

Note that any sensible/common activation function is either linear or of this type. We have shown that in the latter case, together with the identity as activation function, we have an NP-hard reachability problem. Theorem \ref{contin} however only states NP-hardness for this vast set of reachability problems. One might wonder about membership in NP. Our next main result will show that membership in NP and thus NP-completeness is unlikely for many of the activations, because reachability becomes complete for a complexity class conjectured to be much larger than NP.

\begin{definition} cf.\cite{Stefan}
The problem of deciding whether a system of polynomial equations with integer coefficients is solvable over $\mathbb R$ is called the \emph{Existential Theory of the Reals} ETR. The complexity class $\exists\mathbb R$ is defined as the set of all decision problems that reduce to ETR in polynomial time. A problem is called $\exists\mathbb R$-complete if it is in $\exists\mathbb R$ and ETR reduces to the problem in polynomial time.
\end{definition}

It was shown by Canny\cite{canny} that ETR is in PSPACE and it is easily seen that ETR is NP-hard. These are currently the best known bounds and it is widely believed that NP$\subsetneq\exists\mathbb R\subsetneq$PSPACE.

ETR can be formulated as $\CSP(\mathbb{R},E)$, where $E$ is the set of all polynomial relations $R_p:=\{x\in\mathbb{R}^n \mid p(x)\nabla0, \nabla\in\{=,<,\leq\},p\in\mathbb{Z}[x_1,...,x_n]\}$ and inequalities. Note that $E$ does not have finite signature any more, 
this issue has to be resolved before talking about algorithms and complexities of the CSP, for the encoding into Turing Machines is only possible for finite sets of relations. However, $E$ is a first-order reduct of $(\mathbb{R};1,+,\cdot)$, any integer polynomial can be described by the integers, addition, multiplication and logical combinations of these like we did for LP in Example \ref{lpffeaseq}. It thus suffices to represent the relations by their first-order definition. The integer coefficients can be assumed to be encoded in binary by the same ongoing that we used in Example \ref{lpffeaseq}.


The following Theorem will provide us with \NNR\ problems that are hard or even complete for $\exists\mathbb R$. Due to its length, the proof is postponed to the Appendix.
\begin{theorem}\label{alletr}
a) \NNR$(id,f)$ is $\exists\mathbb R$-complete for $f$ any non-linear polynomial and $\exists\mathbb R$-hard for $f$ any function that coincides with a non-linear polynomial on an interval.\\
b) \NNR$(id,f)$ is $\exists\mathbb R$-hard in the following cases: 

i) for any function $f$ that allows to pp-define a function that coincides with the exponential function on an interval, especially the exponential function itself, $ELU(x):=\left\{\begin{matrix}
x \ \ \ \ \ \ & if \ x\geq0\\
\lambda (e^{x}-1) \ \ \ \ & if \ x\leq 0
\end{matrix}\right.$ where $ \lambda\in\mathbb Q$, and $f(x)=e^{-\vert x\vert}$


ii) for the Gaussian function $f(x)=e^{-x^2}$

iii) for the arctan function $f(x)=arctan(x)$ and

iv) for the Gallant-White cosine-smasher $f (x) = \left\{
\begin{array}{ll}
0 & x \leq -\frac\pi2 \\
\frac{1+cos(\frac32\pi)}{2} & x\in[-\frac\pi2,\frac\pi2]  \\
0 & x \geq \frac\pi2 \\
\end{array}
\right. $
.

\end{theorem}
\begin{proof}
See Appendix.\hfill$\blacksquare$
\end{proof}

\begin{remark}
Note that \NNR$(id,e^{(\cdot)})$ is equivalent to $\CSP(\mathbb R; R_+,R_\cdot,1,e^{(\cdot)})$, also known as Tarski's exponential function problem. It is not even known whether this problem is decidable or not (cf.\cite{Macintyre}).
\end{remark}

\section{Network Equivalence and Verification of Interval Property}\label{NEVIP}
In this section, we study the complexity of the remaining decision problems \VIP\ and \NE\ introduced in Definition \ref{decision}. We will see that in a lot of cases these problems are essentially the same.

\begin{theorem}\label{NetEqu}  Let $F$ be a set of activation functions such that $sign, id \in F$.

a) \NE$(F)$ one-one reduces to the complement of \NNR$(F)$ in linear time. Consequently, \NE(id) is in P.

b) \NNR$(F)$ one-one reduces to the complement of \NE$(F)$ in linear time. Consequently, \NE(ReLU) is co-NP-complete and \NE($f$) is co-NP-hard for any non-linear $f$ that is continuous on an interval.

c) \NE \ truth-table reduces to \NE\ with just one output dimension in linear time independent of the set of activation functions.

The same holds for Heaviside or a similar step function instead of sign, id can independently be replaced by ReLU.
\end{theorem}
\begin{proof}
See Appendix.\hfill$\blacksquare$
\end{proof}

\begin{theorem}\label{VIPREACH} Let $F$ be a set of activation functions.

a) \VIP$(F)$ truth-table reduces to the complement of \NNR$(F)$ in linear time. Consequently, \VIP(id) is in P and \VIP(ReLU) is in co-NP.

b) Let $F$ contain at least one among the functions $H$ (the Heaviside function), $sign$ or $ReLU$. Then \NNR$(F)$ one-one reduces to the complement of \VIP$(F)$ in linear time.

c) \VIP$(F)$ truth-table reduces to \VIP$(F)$ with just one output condition in linear time, meaning the output constraint is just one strict inequality.
\end{theorem}

\begin{proof}
See Appendix.\hfill$\blacksquare$
\end{proof}

\begin{theorem}\label{VIPNE} Let $F$ be a set of activation functions containing $id$ or $ReLU$ and $H$ or $sign$.

a) \NE$(F)$ one-one reduces to \VIP$(F)$ in linear time.

b) \VIP$(F)$ one-one reduces to \NE$(F)$ in linear time.
\end{theorem}

\begin{proof} 
See Appendix.\hfill$\blacksquare$
\end{proof}

\section{Conclusion and Further Questions}
We examined the computational complexity of the reachability problem for neural networks in dependence of the activation functions used. We provided conditions for includedness and hardness for NP, translated a dichotomy result for certain CSPs into the language of reachability problems, and showed ETR-hardness of the reachability problems for several typical activation functions. We also showed that \NE\ and \VIP\ are in many cases the same problem which is essentially "co-\NNR". Further open questions are:

\begin{description}
\item[1.)] Do there exist activation functions that are not semilinear but still lead to a reachability problem in NP?
\item[2.)] Can the reachability problem for the sigmoid function $f(x) = 1/(1 + e^{-x})$, one oft the most frequently used activations, be classified any better than just as NP-hard? Can it be related to ETR?
\item[3.)] Can the discussed ETR-hard problems be classified with respect to (potentially) larger complexity classes such as PSPACE, EXP-Time, decidable,...?
\item[4.)] Can reductions between \VIP, \NE\ and \NNR\ be found that do not rely on certain functions to be included in the set of activations?
\item[5.)] How do these problems behave when the deciding algorithm is considered as a computation model with reals as entities? For example of which complexity are the respective problems in a model of real computations like the Blum-Shub-Smale model\cite{BSS}? What if the underlying neural net may have any real weights and biases instead of just rational ones?
\end{description}
\textbf{Acknowledgment}: I want to thank Klaus Meer for helpful discussion.

\bibliography{literatur}

\bibliographystyle{plain}

\newpage
\section{Appendix}
\subsection{Proof of Theorem \ref{alletr}}

The proof requires the following new concept:

\begin{definition}
Let $\tau$ and $\sigma$ be finite relational signatures, $\mathcal{A}$ a $\tau$-structure and $\mathcal{B}$ a $\sigma$-structure. A \emph{primitive positive interpretation} $I$ of $\mathcal{B}$ in $\mathcal{A}$ consists of:

\begin{description}
\item[$\bullet$] A natural number $d\in\mathbb{N}$ called the \emph{dimension},
\item[$\bullet$] a pp - $\tau$ - formula $\delta_I(x_1,...,x_d)$  called the \emph{domain formula},
\item[$\bullet$] for each atomic $\sigma$ - formula $\phi(y_1,...,y_k)$, a pp - $\tau$ - formula $\phi_I(\bar{x}_1,...,\bar{x}_k)$ called the \emph{defining formulas} and
\item[$\bullet$] a surjective map $h\colon D \rightarrow B$ called the \emph{coordinate map},  where\\
$D\coloneqq \{(a_1,...,a_d)\mid a_1,...,a_d\in A , \mathcal{A}\models \delta_I(a_1,...,a_d)\}$, $A$ the domain of $\mathcal{A}$ and $B$ the domain of $\mathcal{B}$
\end{description}
so that the following holds for every atomic $\sigma$-Formula $\phi$ and all $\bar{a}_1,...,\bar{a}_k \in D$:

\[\mathcal{A}\models \phi(h(\bar{a}_1),...,h(\bar{a}_k))  \Leftrightarrow  \mathcal{A}\models \phi_I(\bar{a}_1,...,\bar{a}_k)\]

\end{definition}

It was shown in \cite{bodirsky} (Theorem 3.1.4), that if such a pp-interpretation of $\mathcal{B}$ in $\mathcal{A}$ exists, then there is a polynomial-time reduction from $\CSP(\mathcal{B})$ to $\CSP(\mathcal{A})$.

\begin{lemma}\label{R+}
ETR is polynomial-time (in fact linear-time) equivalent to 

a) ETR on the domain $\mathbb R^+:=\{x\in\mathbb R \mid x\geq0\}$, denoted by ET$(\mathbb R^+)$

b) ETR on the domain $\mathbb R\cap(0,\frac1n]$ for any $n\in\mathbb N$ denoted by ET$(\mathbb R\cap(0,\frac1n])$.

\end{lemma}

\begin{proof} a) We can easily reduce an instance of ET($\mathbb R^+$) into an instance of ETR by additionally demanding that its variables have to be positive:
\[x\geq 0\Leftrightarrow \exists y\in\mathbb R : y^2=x.\]
For the reverse direction, pp-interpret $(\mathbb R,+,\cdot)$ 2-dimensional ($d=2$) in $(\mathbb R^+,+,\cdot)$ as follows:
The domain of the interpretation structure is the set of pairs 
\[D=\{(a,b)\mid a,b\in\mathbb R^+ , a\cdot b=0\}, \text{ so } \delta_I(a,b)\Leftrightarrow a\cdot b=0.\] For every variable $x$ in the ETR-instance, the interpreting pair $(a,b)$ consists of its positive part $a=max(x,0)$ and its negative part $b=max(-x,0)$, thus at least one of them must be zero, the coordinate map is therefore 
\[h(a,b)=a-b.\]
1 is interpreted by 
\[1_I(a,b)\Leftrightarrow a=1\]
The sum $(a,b)+_I(c,d)=(e,f)\Leftrightarrow a+c=e\land b+d=f$ could violate the rule that the product of the parts has to be zero, meaning $(a+c,b+d)$ is in general not in the domain $D$. Repair it via
\[(a,b)+_I(c,d)=(e,f) :\Leftrightarrow\exists\varepsilon : a+c-\varepsilon=e \land b+d-\varepsilon=f.\]
Finally we define multiplication employing the same idea:
\[(a,b)\cdot_I(c,d)=(e,f) :\Leftrightarrow\exists\varepsilon : ac+bd-\varepsilon=e \land ad+bc-\varepsilon=f.\]

b) We can again translate an instance of ET($\mathbb R\cap(0,\frac1n]$) into an instance of ETR by requiring its variables to be in $(0,\frac1n]$.

For the reverse direction, we first use part a) so we can work with $\mathbb R^+$. Note that we can replace $\mathbb R^+$ by $\mathbb R\cap[n,\infty)$ by a 1-dimensional pp-interpretation $I$ with domain
\[D=\{a\mid \}, \text{ so } \delta_I(a)\Leftrightarrow T\]
and \[h(a)=a-n,\]
so the defining formula for 1 is 
\[1_I(a)\Leftrightarrow a=n+1\] and for addition
\[a+_Ib=c\Leftrightarrow a+b=c+n.\]
For multiplication we have 
\[a\cdot_Ib=c\Leftrightarrow (a-n)\cdot (b-n)=(c-n).\]

The final domain is the set $D=(0,\frac1n]$, we map every $x\geq n$ to its inverse:
\[h_J(x)=\frac1x\]
This allows us to interpret multiplication:
\[a\cdot b=c \Leftrightarrow h_J(a)\cdot h_J(b)=h_J(c)\]
Addition is a bit more complicated,
\[a+b=c\Leftrightarrow \frac1{h_J(a)}+\frac1{h_J(b)}=\frac1{h_J(c)}\]
is not sufficient, for $\frac1{h_J(a)},\frac1{h_J(b)},\frac1{h_J(c)}\notin(0,\frac1n]$. We additionally need
\begin{align*}
\frac1{h_J(a)}+\frac1{h_J(b)}=\frac1{h_J(c)}\Leftrightarrow& \frac{h_J(b)+h_J(a)}{h_J(a)\cdot h_J(b)}=\frac1{h_J(c)}\\
\Leftrightarrow& (h_J(b)+h_J(a))\cdot h_J(c)=h_J(a)\cdot h_J(b)\\
\Leftrightarrow& h_J(b)\cdot h_J(c)+h_J(a)\cdot h_J(c)=h_J(a)\cdot h_J(b)\\
\end{align*}
All of the multiplications in the last step give us values that are smaller than $\frac1n$, because we multiply values $\varphi(a),\varphi(b),\varphi(c)$ that are smaller than $\frac1n$. The addition on the left side could give us a result bigger than $\frac1n$, but only if the initial equation was incorrect (in which case we want the last equation to be unsolvable).\hfill$\blacksquare$
\end{proof}

We prove the statements of Theorem \ref{alletr} separately:


\begin{proof}[Theorem \ref{alletr}, a)] For $p$ a polynomial, 
\NNR$(id,p)$ obviously reduces to ETR, for $p$ can be pp-defined by addition, multiplication and rational constants.

For completeness, we want to pp-define $f(x)=x^2$, let therefore $p(x)=\sum\limits_{i=0}^na_ix^i$ with $a_n=1 , n\geq 2$. Then $p(x+k)$ is pp-definable in $(\mathbb R; \leq, +, 1, p)$ for every $k\in\mathbb N$, where we interpret the function $p$ as a binary relation again. The idea is now to find a linear combination of these $p(x+k)$ that is equal to $x^2$ by inductively pp-defining polynomials of smaller degrees. We have
\begin{align*}
p(x+k)&=\sum\limits_{i=0}^na_i(x+k)^i\\
&=\sum\limits_{i=0}^na_i\sum\limits_{j=0}^i\binom ijx^jk^{i-j}\\
&=\sum\limits_{i=0}^n\sum\limits_{j=0}^ia_i\binom ijx^jk^{i-j}\\
&=\sum\limits_{j=0}^n\sum\limits_{i=j}^na_i\binom ijx^jk^{i-j}\\
&=\sum\limits_{j=0}^nx^j\sum\limits_{i=j}^na_i\binom ijk^{i-j}\\
\end{align*}

Note that for every $x^j$, the dominant term for $k\rightarrow \infty$ is the last one ($i=n$, namely $a_n\binom njk^{n-j}=\binom njk^{n-j}$), note that it is never zero. In every polynomial $p_{1,k}:=p(x+k) , k\in\mathbb N$, the highest degree monomial is always $x^n$ with coefficient 1. The polynomials $p_{2,k}(x):=p_{1,k}(x)-p(x) , k\in\mathbb N$ have no $x^n$ term, they are at most of degree $n-1$. The $x^{n-1}$ term does not vanish for all but finitely many of the $p_{2,k}(x)$, since from every $p_{1,k}$ we subtracted the same polynomial, but the coefficients of the $x^{n-1}$ terms of $p_{1,k}$ grow linear in $k$. Take the first such polynomial with an $x^{n-1}$-term and divide it by its leading coefficient.

Repeat this step until a constant polynomial is reached. Now we have constructed for every $i\in\{0,...,n\}$ a polynomial that is exactly of degree $i$, so they form a basis of $\langle 1,x,x^2,...,x^n\rangle$. We can therefore pp-define $f(x)=x^2\in\langle 1,x,x^2,...,x^n\rangle$. It remains to show that $\CSP(\mathbb R; \leq, R_+,R_{=1}, (\cdot)^2)$ is ETR-hard. Multiplication can be pp-defined in $(\mathbb R; \leq, R_+,R_{=1}, (\cdot)^2)$ via
\[a\cdot b=\frac{(a+b)^2-a^2-b^2}{2}\] 
This concludes the proof, for multiplication was the only relation missing for ETR and the proof of Lemma \ref{R+} shows that it suffices to have multiplication on an arbitrary interval.\hfill$\blacksquare$
\end{proof}

\begin{proof}[Theorem \ref{alletr}, b),i)]
It suffices by Lemma \ref{R+} to reduce ET($\mathbb R^+$) to\\
\NNR$(id,e^{(\cdot)})$, this is possible in linear time:

Positivity, additions and rational constants can be encoded in \NNR$(id)$ as before. It remains to define multiplication, by Theorem \ref{addfunction} it suffices to pp-define it in $(\mathbb R; \leq,R_+,R_{=1}, e^{(\cdot)})$:
\begin{align*}
x\cdot y=z\Leftrightarrow \ & (x+1)\cdot(y+1)=(z+1)+x+y\\
\Leftrightarrow \ & log(x+1)+log(y+1)=log((z+1)+x+y)
\end{align*}
The logarithm is pp-definable as the inverse of the exponential function.

To show that the exponential function on an arbitrary interval $[a,b]$ already suffices, observe that it enables the multiplication on the interval $[e^a-1,e^b-1]$
\[R_{\bullet,[e^a-1,e^b-1]}:=\{x,y,z \mid x\cdot y=z \land x,y\in[e^a-1,e^b-1]\}.\]
It remains to show that multiplication $R_{\bullet,[e^a-1,e^b-1]}$ on an interval suffices to pp-define multiplication on $(0,\frac1n]$. Choose $e^a-1<\alpha<\beta<e^b-1$ with $\alpha,\beta\in\mathbb Q$ and choose $n$ so that $\frac1n\leq \beta-\alpha$. For $x,y,z\in(0,\frac1n]$ we have
\begin{align*}
x\cdot y=z&\Leftrightarrow (x+\alpha)\cdot (y+\alpha)=z+\alpha x+\alpha y+\alpha^2\\
&\Leftrightarrow (x+\alpha)\cdot (y+\alpha)-z-\alpha^2=\alpha x+\alpha y
\end{align*}
The multiplication $(x+\alpha)\cdot (y+\alpha)$ can be performed, because both values are in $[\alpha,\beta]$. The subtraction of the rational constant $\alpha^2$ and variable $z$ are possible as well. Thus, the value $t=(x+\alpha)\cdot (y+\alpha)-z-\alpha^2$ is pp-definable in $(\mathbb R^+; \leq,R_+,R_{=1}, e^{(\cdot)}\vert_{[a,b]})$. Observe that 
\begin{align*}
x\cdot y=z&\Leftrightarrow (x+\alpha)\cdot (y+\alpha)-z-\alpha^2=\alpha x+\alpha y\\
&\Leftrightarrow t=\alpha x+\alpha y\\
&\Leftrightarrow \frac t2+\alpha^2=\frac{\alpha x+\alpha y}{2}+\alpha^2\\
&\Leftrightarrow \frac t2+\alpha^2=\alpha(\frac{x+y}{2}+\alpha)\\
\end{align*}
For division by 2, demand a new variable $t'$ to fulfill $t=t'+t'$, adding $x,y$ and $\alpha$ is also possible, as well as the multiplication of $\alpha $ and $\frac{x+y}{2}+\alpha$, for both are in $[\alpha,\beta]$. This proves the first statement.

It follows that also the activation functions $ELU(x):=\left\{\begin{array}{ll}
x \ \ \ \ & if \ x\geq0 \\
\lambda (e^{x}-1) \ \ \ & if \ x\leq 0\\
\end{array}\right.$ and the function $f(x)=e^{-\vert x\vert}$ both result in ETR-hard \NNR-problems.\hfill$\blacksquare$
\end{proof}



\begin{proof}[Theorem \ref{alletr}, b),ii)]
We have that $f(2x)=e^{-(2x)^2}=e^{-4x^2}=(e^{-x^2})^4=f(x)^4$.
This means that we can pp-define the function 
\[g:(0,\infty)\rightarrow \mathbb R,  g(x)=x^4\vert_{(1,\infty)}\] by $g(x)=f(2\cdot f^{-1}(x))$. By $f^{-1}$ we mean any of the preimages of $x$ under $f$. Such a preimage can be pp-defined and the value of $g(x)$ is the same for both preimages. Now $\exists\mathbb R$-hardness follows by Lemma \ref{R+} and Theorem \ref{alletr} a).\hfill$\blacksquare$
\end{proof}

\begin{proof}[Theorem \ref{alletr}, b),iii)]
It suffices again to pp-define a non-linear polynomial on an interval. To achieve this, recall that $arccot(x)=\frac\pi2-\arctan(x)$ and that $\arctan(1)=\frac\pi4$ is pp-definable, so we can easily switch to $arccot$. Next, recall
\[arccot(x)+arccot(y)=arccot(\frac{xy-1}{x+y})\]
for $x,y>0$. It follows that\begin{align*}
arccot(x)+arccot(4x^3+3x)&=arccot(\frac{4x^4+3x^2-1}{4x^3+4x})\\
&=arccot(\frac{(4x^2-1)(x^2+1)}{4x(x^2+1)})\\
&=arccot(\frac{(2x)^2-1}{2x+2x})\\
&=2arccot(2x)\\
\end{align*}The non-linear polynomial $f(x)=4x^3+3x$ can therefore be pp-defined by 
\[f(x)=cot(2arccot(2x)-arccot(x))\]
which concludes the proof.\hfill$\blacksquare$
\end{proof}

\begin{proof}[Theorem \ref{alletr}, b),iv)]
It suffices again to pp-define a non-linear polynomial on an interval. To achieve this, recall that $cos(2x)=2cos^2(x)-1$

The non-linear polynomial $f(x)=2x^2-1$ can therefore be pp-defined by 
\[f(x)=\frac{cos(2\cdot arccos(x))-1}{2}\]
which concludes the proof.\hfill$\blacksquare$
\end{proof}

\subsection{Proofs of Results in Chapter \ref{NEVIP}}
\begin{proof} [Theorem \ref{NetEqu}]a) If the networks $N_1$ and $N_2$ from the \NE-instance have different numbers of layers, we "copy" the output of the "flatter" one by the identity function until both have the same size. The idea is now to construct a network $N$ that has the same input dimension as $N_1$ and $N_2$ and computes 0 on an input $x$ if and only if $y_1=N_1(x)=N_2(x)=y_2$ and $\pm1$ otherwise.

To achieve this we put both networks $N_1$ and $N_2$ into one network $N$ by contracting the corresponding input nodes and add one layer that computes the difference between the outputs of $N_1$ and $N_2$. More precisely, it computes for every output dimension $i$ the signum of the difference $y_i=sign(y_{1,i}-y_{2,i})$. We have equivalence of $N_1$ and $N_2$ iff all $y_i$ are always 0. We then add another layer that computes the signum of their weighted sum $y=sign(\sum\limits_{i=1}^m 2^i\cdot y_i)$, which is 0 iff all $y_i$ are 0. In the last two layers we compute $z=sign(sign(y+1)-sign(y-1))$ which is 1 iff $y=\pm1$ and 0 iff $y=0$. $N$ together with the output specification $z\geq\frac12$ is now an \NNR$(F)$-instance that is true iff $N_1$ and $N_2$ are not equivalent. 

b) We construct two new networks $ N_1, N_2$ depending on the network $N$ and the specifications that form the instance of \NNR$(F)$. The idea is that $N_1(x)=0$ iff both input and output specification of $N$ are fulfilled and -1 otherwise and that $N_2$ computes $-1$ independent of the input.\\
The specification is a conjunction of equations of the form 
\[(E_k) \ \ \sum\limits_{i=1}^n a_{i+k}v_{i,k}\geq a_k\] for rational $a_{i+k},a_k$ and the $v_{i,k}$ either all in the input or all in the output layer. For each such equation $E_k$, we introduce two connected $sign$-nodes that compute $\lambda_k=sign(sign(\sum\limits_{i=1}^n a_{i+k}v_{i,k}- a_k)+\frac12)$. For every input, the result $\lambda_k$ is 1 if and only if the condition $E_k$ is fulfilled and $-1$ otherwise.  
We introduce another $sign$-node that computes $\lambda=sign(\sum\limits_{k=1}^m \lambda_k-m)$, the result is 0 if and only if all specifications are fulfilled and -1 otherwise, $\lambda$ is the overall network's output.

The second network $ N_2$ that we construct has the same number of input-nodes $v_1,...,v_\ell$ as $N$ and computes $sign(sign(v_1)-2)$ which is $-1$ independent of the input. This means that the \NNR$(F)$-instance $N$ is solvable if and only if the \NE$(F)$-instance $( N_1, N_2)$ returns false, implying $ N_1$ and $ N_2$ do not compute the same function.

c) To truth-table reduce \NE \ on \NE\ with just one output dimension, we ask for every output dimension if both networks always compute the same value. If that is always the case, we accept, otherwise we reject.\hfill$\blacksquare$
\end{proof}

\begin{proof}[Theorem \ref{VIPREACH}] a) To reduce \VIP\ to the complement of \NNR, we ask for every output constraint that describes $B$ if its negation is satisfiable. If that is the case at least once, we reject, otherwise we accept. Note that this way every strict inequality describing an open half space becomes a non strict inequality describing the complement of an open half space, which is a closed half space.

b) For both $H$ and $sign$, we construct an instance of VIP equivalent to the negation of an instance $(N,A,B)$ of \NNR$(F)$. Let $H\in F$. The input constraints $A$ stay the same, the network $N'$ is a copy of $N$ with two additional layers following the original output layer. The first one contains for each output constraint $\sum\limits_{i=1}^k a_ix_i\leq b$ in $B$ an $H$-node computing $\bar a = H(b-\sum\limits_{i=1}^k a_ix_i)$. The second added layer is the output layer of $N'$ and contains only one $H$-node computing $ \hat a = H(\sum\limits_{i=1}^k \bar a_i-n)$. The output $ \hat a$ of $N'$ is 1 if the output of $N$ is in $B$ and 0 otherwise. We therefore choose the interval $[-\frac12,\frac12]$ as output constraint of \VIP. Now
\[\text{\NNR}(F)(N,A,B)\Leftrightarrow\lnot \text{\VIP}(F)(N',A,(-\frac12,\frac12))\]
The proof for sign is analogous.

For $ReLU\in F$, we choose again the same input constraints $A$ and the network $N'$ is a copy of $N$ with two additional layers after the previous output layer. This time, the first layer contains for each output constraint $\sum\limits_{i=1}^k a_ix_i\leq b$ in $B$ a $ReLU$-node computing $\bar a = ReLU(\sum\limits_{i=1}^k a_ix_i-b)$ and the second added layer (again the output layer) contains only one $ReLU$-node computing $\hat a = ReLU(\sum\limits_{i=1}^k \bar a_i)$. The constraints are simultaneously fulfilled if and only if $ \hat a =0$, so 
\[\text{\NNR}(F)(N,A,B)\Leftrightarrow\lnot \text{\VIP}(N',A,(0,\infty))\]

c) To truth-table reduce \VIP \ to \VIP\ with just one condition, we ask for every output constraint that describes $B$ if it is always satisfied. If that is the case for every condition, we accept, otherwise we reject.\hfill$\blacksquare$
\end{proof}

\begin{proof}[Theorem \ref{VIPNE}] a) Follows by Theorem \ref{NetEqu} a) and Theorem \ref{VIPREACH}.

b) Let $(N,A,B)$ be a \VIP$(F)$-instance and assume that $id,sign\in F$, the proof for $ReLU$ and/or $H$ is analogous. We construct two networks $N_1$ and $N_2$ so that $N_1(x)=1$ when $x\in A,N(x)\notin B$ and $N_1(x)=-1$ otherwise, and $N_2(x)=-1$ independent of $x$. $N_2$ is the network $N$ with the following nodes attached: For every input condition $\sum\limits_{i=1}^n a_{i+k}v_{i,k}\geq a_k$ we add two nodes computing $b_k=sign(sign(a_k-\sum\limits_{i=1}^n a_{i+k}v_{i,k})+\frac12)$. We have $b_k=1$ if the condition holds and $b_k=-1$ otherwise. The next node computes $b=sign(\sum\limits_{k=1}^\ell \frac{b_k}{\ell}-1)$, which is 0 if all input conditions are simultaneously fulfilled and $-1$ otherwise. For the output constraints, we apply a similar construction to the output nodes to obtain a node computing $c=-1$ if the conditions hold and $c=1$ otherwise. The overall output of $N_1$ is $d=sign(b+c-\frac12)$, which is 1 iff $x\in A,N(x)\notin B$ and $-1$ otherwise. $N_2$ computes $sign(sign(x_i)-2)$ for any input $x_i$, so it is always $-1$. This means that \VIP$(N,A,B)$ does not hold, meaning there is an input in $A$ that is not mapped to $B$, iff $N_1$ and $N_2$ are not equivalent.\hfill$\blacksquare$
\end{proof}


\end{document}